\documentclass[12pt,preprint]{aastex}

\usepackage{graphicx}

\slugcomment{Astrophysical Journal Letters}

\shorttitle{P/2013 P5}
\shortauthors{Jewitt et al.}

\begin{document}

\title{The Extraordinary Multi-Tailed Main-Belt Comet P/2013 P5
}
\author{David Jewitt$^{1,2}$, Jessica Agarwal$^3$, Harold Weaver$^4$,  Max Mutchler$^5$ and Stephen Larson$^6$
}
\affil{$^1$Department of Earth and Space Sciences,
University of California at Los Angeles, \\
595 Charles Young Drive East, 
Los Angeles, CA 90095-1567\\
$^2$Department of Physics and Astronomy,
University of California at Los Angeles, \\
430 Portola Plaza, Box 951547,
Los Angeles, CA 90095-1547\\
$^3$ Max Planck Institute for Solar System Research, Max-Planck-Str. 2, 37191 Katlenburg-Lindau, Germany \\
$^4$ The Johns Hopkins University Applied Physics Laboratory, 11100 Johns Hopkins Road, Laurel, Maryland 20723  \\
$^5$ Space Telescope Science Institute, 3700 San Martin Drive, Baltimore, MD 21218 \\
$^6$ Lunar and Planetary Laboratory, University of Arizona, 1629 E. University Blvd.
Tucson AZ 85721-0092 \\
}

\email{jewitt@ucla.edu}

\begin{abstract}
Hubble Space Telescope observations of main-belt comet P/2013 P5 reveal an extraordinary system of six dust tails that distinguish this object from any other.  Observations two weeks apart show dramatic morphological change in the tails while providing no evidence for secular fading of the object as a whole.  Each tail is associated with a unique ejection date, revealing continued, episodic mass loss from the 0.24$\pm$0.04 km radius nucleus over the last five months.   As an inner-belt asteroid and probable Flora family member, the object is likely to be highly metamorphosed and unlikely to contain ice. The protracted period of dust release appears inconsistent with an impact origin, but may be compatible with a body that is losing mass through a rotational instability.  We suggest that P/2013 P5 has been accelerated to breakup speed by radiation torques.  
\end{abstract}

\keywords{minor planets, asteroids: general --- minor planets, asteroids: individual (P/2013 P5) --- comets: general}

\section{Introduction}
P/2013 P5 (hereafter P5) was announced on UT 2013 Aug 27 \citep{2013MPEC....Q...37M}.  Orbiting at the inner edge of the main asteroid belt, with semimajor axis, eccentricity and inclination of 2.189 AU, 0.115 and 5.0\degr, respectively, the Tisserand parameter relative to Jupiter is $T_J$ = 3.66. This is far above the nominal $T_J$ = 3 dividing line that separates dynamical comets from asteroids.  Despite this, P5 displays a tail in the discovery data, suggesting that it has ejected material.  The combination of asteroid-like orbit and comet-like appearance together reveal P5 as an active asteroid (equivalently, a main-belt comet - MBC).  No known dynamical path connects the main-belt to the Kuiper belt or Oort cloud comet reservoirs.  For this reason, the active asteroids are regarded as a distinct class of solar system body.  While some are suspected to contain water ice whose sublimation is responsible for the expulsion of dust (Hsieh and Jewitt 2006), others are impact-produced while, for a majority, the origin is unknown\citep{2012AJ....143...66J}.  

In this brief paper we describe initial high angular resolution images of P5 taken using the Hubble Space Telescope to attempt to determine its basic properties and to establish the cause of the observed mass loss.

\section{Observations} 
We used two orbits of Target-of-Opportunity time (General Observer program number 13475) to observe P5 on UT 2013 Sep 10 and 23, obtaining a total of 12 images with the WFC3 camera \citep{2010AAS...21546322D}.  The 0.04\arcsec~pixels each correspond to about 33 km at the distance of P5, giving a Nyquist-sampled (two-pixel) spatial resolution of about 66 km.  All observations were taken using the F350LP filter. This very broad filter (full-width-at-half-maximum 4758\AA) provides maximum sensitivity to faint sources at the expense of introducing some uncertainty in the transformation to standard astronomical filter sets.  The effective wavelength for a solar-type (G2V) source is 6230\AA.  The observational geometry is summarized in Table \ref{geometry}.

\begin{figure}
\epsscale{0.9}
\begin{center}
\plotone{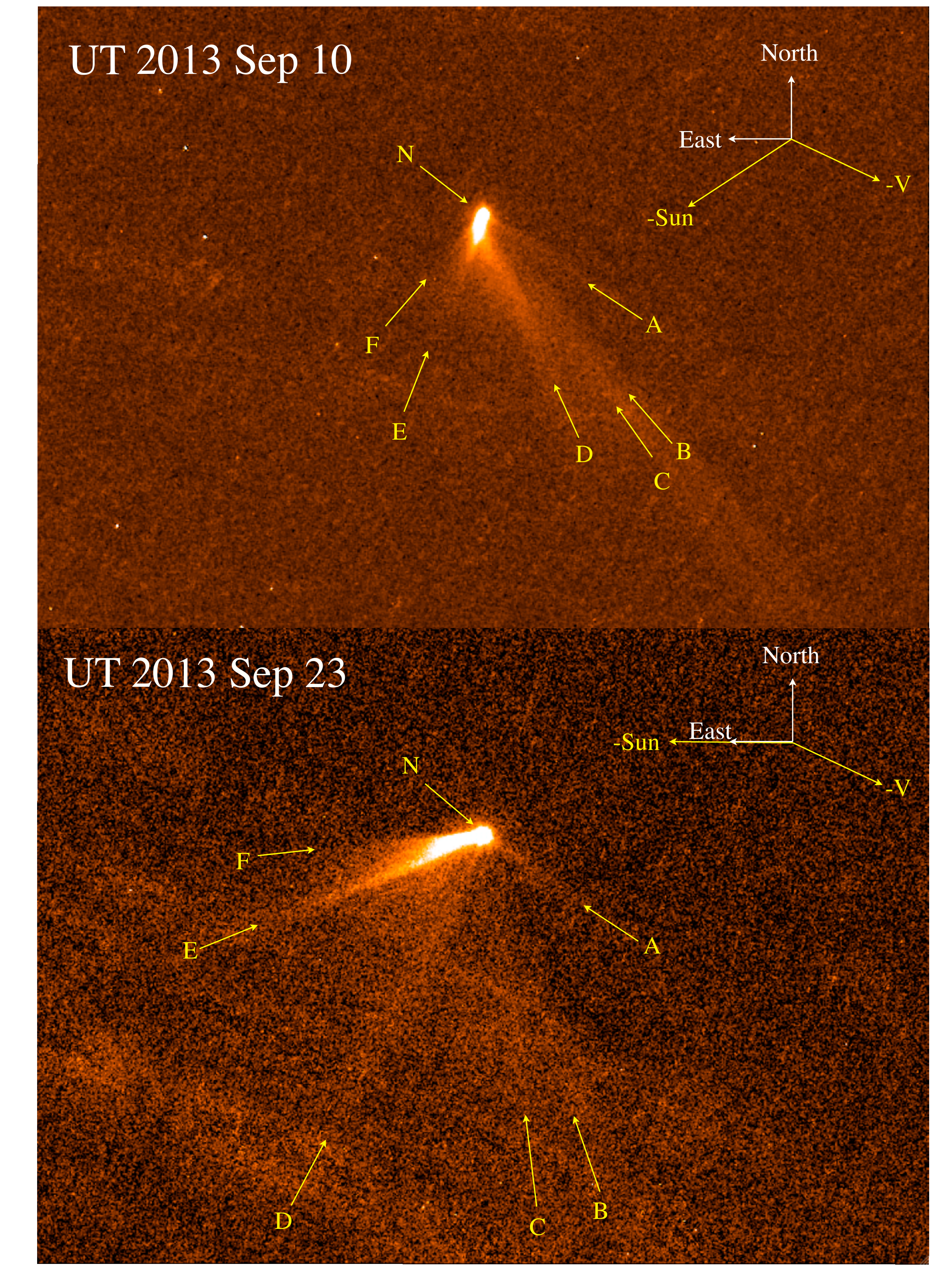}
\caption{Composite images of P/2013 P5 on UT 2013 Sep 10 (upper) and 23 (lower).    Each panel shows a region 28\arcsec~(23,000 km) in width, with cardinal directions as marked.  Letters denote features described in the text.  The projected antisolar and negative velocity vectors are marked.  \label{image}
} 
\end{center} 
\end{figure}

On each orbit we obtained five exposures of 348 s and one of 233 s. The drizzle-combined images for each date are shown in Figure (\ref{image}).  On both dates, the images show a structured, multiple tail system extending $>$25\arcsec~ to the edge of the field of view.   Important morphological features of the object are labeled with letters in Figure (\ref{image}).  The changes in the $\sim$2 week interval between observations are dramatic.  In both images a centrally condensed nucleus, N,  is apparent, but most other features have changed in both brightness and position, making correlations between images difficult.  Identifications of tails A-F are marked in both panels of Figure (\ref{image}).  Additional diffuse features, especially in the data from UT 2013 Sep 23, are imperfectly removed field galaxies that have been swept across the field of view by non-sidereal tracking.  

\subsection{Nucleus}
For photometry of the nucleus, we used apertures 5 pixels (0.2\arcsec) and 25 pixels (1.0\arcsec) in radius, with sky subtraction determined from the median signal computed within a contiguous annulus having outer radius 50 pixels (2.0\arcsec).  We used the HST exposure time calculator to convert the measured count rate into an effective V magnitude, finding that a V = 0 G2V source gives a count rate of 4.72 $\times$10$^{10}$ s$^{-1}$ within a 0.2\arcsec~radius photometry aperture.   The results are summarized in Table (\ref{photometry}).

The smaller aperture gives our best estimate of the nucleus brightness.   We examined the six images from each orbit individually to search for temporal variation that might result from rotation of an irregular nucleus.  No such variation was observed on either date, consistent with a nucleus rotation period that is long compared to the $\sim$40 minute observing window per HST orbit, or a rotation axis that is close to the line of sight, or both.  The mean apparent magnitudes on Sept 10 (V = 20.92$\pm$0.01) and 23 (21.01$\pm$0.01), are very similar, given that the observing geometry changed between the two dates of observation (Table \ref{geometry}).

We compute the absolute magnitude (i.e.~corrected to unit heliocentric and geocentric distances, and to zero phase angle) using the inverse square law and an assumed phase function from \citet{1989aste.conf..524B}.  We use a phase function parameter $g$ = 0.25, consistent with an S-type asteroid spectral classification, since most inner-belt asteroids are S-types.  The phase corrections on Sep.~10 and Sep.~23 are -0.37 and -0.59 mag., respectively.  If we had instead assumed $g$ = 0.15, corresponding to the phase function of a C-type asteroid, the corrections would have been -0.42 and -0.67 mag.   Our lack of knowledge of the phase function renders the derived absolute magnitudes uncertain by 0.05 to 0.1 mag.  Regardless, the estimated values $H_V$ = 18.69 (Sep.~10) and 18.54 (Sep.~23) are very close.  The 0.15 mag.~brightening in $H_V$ could be due to rotational variation of an elongated nucleus between the two dates, to phase function uncertainties or to an increase in the amount of near-nucleus dust.  There is no evidence for fading that might be expected if the dust were ejected impulsively before the first observation.

The absolute magnitude is related to the effective nucleus radius, $r_n$ (in km), and to the geometric albedo, $p_V$, by

\begin{equation}
r_n = \frac{690}{p_V^{1/2}}10^{-H_V/5}
\label{inversesq}
\end{equation}.  

\noindent where $p_V$ is the V-band geometric albedo.    The orbit of P5 is close to the Flora asteroid family (the family center lies near $(a,e,i)$ = (2.254,0.141,5.5)).  The mean geometric albedo of Flora family members is $p_V$ =  0.29$\pm$0.09 \citep{2013ApJ...770....7M}, which we take as the albedo of P5 pending a future measurement.   Substitution into Equation (\ref{inversesq}) then gives $r_n$ = 0.24$\pm$0.04 km, where the uncertainty reflects only the $\sim$30\% uncertainty in $p_V$.  The Floras are thought to be a principal source of near-Earth asteroids and meteorites, and a particular association with the LL chondrites has been claimed (Dunn et al.~2013).  The average mean density of LL chondrites is $\rho_{LL}$ = 3300$\pm$200 kg m$^{-3}$ \citep{2000M&PS...35.1203W,2008ChEG...68....1C}.  With this density and radius, the approximate gravitational escape speed from the nucleus is $V_e$ = 0.3 m s$^{-1}$.  Strictly, both $r_n$ and $V_e$ are upper limits to the true values, since even the small aperture photometry must include some near-nucleus dust contamination for which we have made no correction.

\subsection{Tails}
The position angles of the tails are summarized in Table (\ref{angles}), together with best estimates of their uncertainties based on repeated measurements at different positions along the tails.  Tail A changed least between the visits,  in terms of its direction, length and  brightness.  The tail A position angle is closest to, but significantly different from, the projected orbital velocity vector, marked -V in the figure.  This observation suggests that A contains larger, slower, possibly older particles than those in the other tails.  Tails B, C and D splay apart between Sep 10 and 23, and also fade in surface brightness.  Their large angular motion is presumably related to the $\sim$36\degr~change in the projected antisolar position angle, $\theta_{\odot}$ (Table \ref{geometry}), but the tails rotate by larger angles and are not antisolar.  The fading might suggest dissipation of the tails under the action of radiation pressure.  While B, C and D fade,  tail E grows in length and brightens between the two visits indicating that fresh material is being supplied to it.  

\begin{figure}
\epsscale{0.85}
\begin{center}
\plotone{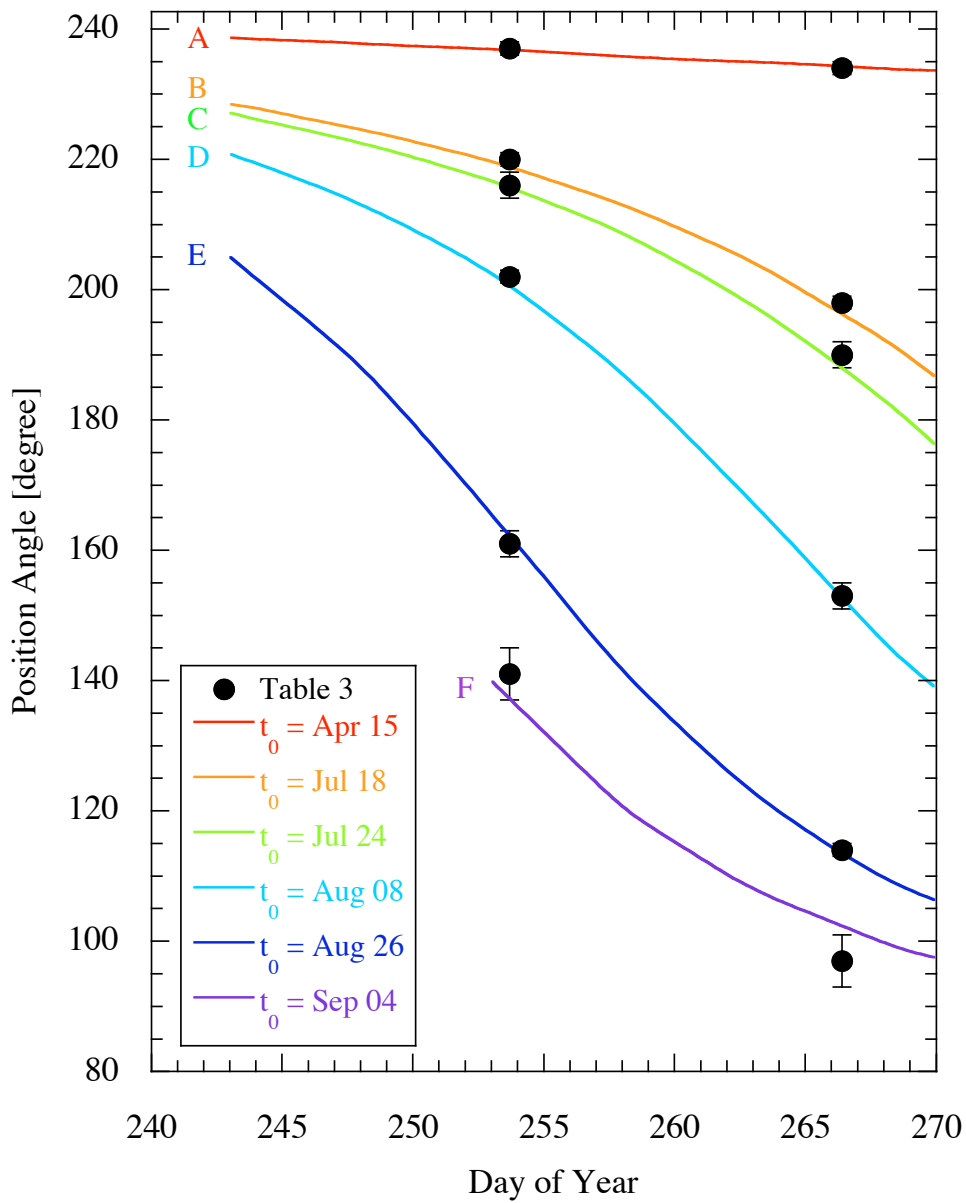}
\caption{Position angles of the tails from Table \ref{angles} (black circles) and calculated synchrones (solid lines) as functions of the date of observation.  For any given tail, the position angles measured on the two observations dates are consistent with the same date of dust ejection (color coded), with an uncertainty of less than a day. The synchrone initiation times, $t_0$, are listed. \label{synchrones}
} 
\end{center} 
\end{figure}

We calculated the positions of particles of a wide range of sizes and released over a wide range of dates, assuming that the initial velocity of the particles is negligible compared to the combined action of solar radiation pressure and gravity \citep{1968ApJ...154..327F}. All particles ejected at a given time lie on a straight line emerging from the nucleus (a synchrone), the orientation of which (described by the position angle) is diagnostic of the date of ejection. For any given observation date, there is a unique relation between position angle and ejection time.  We find, for a given tail, an excellent agreement between the ejection date derived from the observations on Sep 10 and the ejection date derived from the Sep 23 data (Figure \ref{synchrones} and Table \ref{angles}).  This result shows that the individual tails follow synchrones and that splaying of the tails between the two panels of Figure (\ref{image}) is caused by projection effects, not by evolutionary changes in the tails.  The synchrone models indeed show that tail A is the oldest, with ejection on April 15$\pm$2.  The youngest is tail E, with ejection occurring only days before the Sep 10 observation.  We speculate that the ejection of bright tail D on Aug 09 may have been responsible for the discovery of P5 on Aug 18.  From the lengths of the tails and their best-fit synchrone ages, we estimate particle sizes up to $\sim$10 $\mu$m to $\sim$100 $\mu$m, with the largest particles being those in tail A.

The complicated morphology and low surface brightness of P5 make photometry of the dust very difficult.  We used large circular apertures 4.0\arcsec~and 6.0\arcsec~in radius, with sky subtraction from a surrounding annulus extending to 12.0\arcsec, to measure the integrated light from the dust.  The principal uncertainties on the large aperture magnitudes are systematic and difficult to quantify, but are at least several tenths of a magnitude. The total magnitudes (Table \ref{photometry}) provide an estimate of the total dust cross-section in P5.   On both dates, $\Delta V = V_{0.2}$ - $V_{6.0} \sim$ 2.8 mag., showing no relative fading between the two. If the dust in the tail has the same albedo and phase function as the nucleus, then the dust cross-section can be estimated from $C_d$ = $\pi r_n^2 10^{0.4\Delta V}$.  We find $C_d$ = 2.7$\pm$0.4 km$^2$. 

The total dust mass, $M_d$, and cross-section of an optically-thin assemblage of spherical particles are related by $M_d = 4/3 \rho \overline{a} C_d$, where $\rho$ is the material density and $\overline{a}$ is the cross-section weighted mean grain size.  With mean grain size $\overline{a}$ = 10 $\mu$m to 100 $\mu$m and density $\rho$ = 3300 kg m$^{-3}$, for instance, the  dust mass implied by $C_d$ is $M_d \sim$ 10$^5$ to 10$^6$ kg, very small compared to the mass of the nucleus, assuming the same density.


\section{Discussion}
Processes invoked to explain mass loss from asteroids include sublimation of near surface ice, electrostatic levitation of dust, impact, and rotational instability \citep{2012AJ....143...66J}.  
The orbit of P5 lies near the inner edge of the asteroid belt, in the vicinity of the Flora family of S-type asteroids.  These objects have been associated with the LL chondrites, which themselves reflect metamorphism to temperatures $\sim$800\degr C to 960\degr C \citep{2000P&SS...48..887K}.  As such, P5 is an unlikely carrier of water ice, and sublimation is unlikely to account for the observed activity.  Neither is it likely that P5 could be a comet captured from the Kuiper belt or Oort cloud comet reservoirs; numerical simulations show that such capture is effectively impossible in the modern solar system \citep{2002Icar..159..358F}.   Impact is another potential source of dust in the asteroid belt.  However, the five month age spread of the dust tails  (Table \ref{angles}) and the absence of fading in the photometry (Table \ref{photometry}) both argue strongly against impact as a plausible explanation for the activity in P5.   

The surviving hypothesis is that P5 is a body showing rotational mass-shedding, presumably from torques imposed by solar radiation.    Interestingly, it has been suggested that the spins of Flora family members show statistical evidence for the action of radiation torques \citep{2013A&A...551A.102K} .  Rotational re-shaping and breakup under radiation torques are two of the most interesting subjects in asteroid science \citep[c.f.][]{2011Icar..214..622M,2012Icar..218..876S}. Unfortunately, the expected observational signature of a rotationally disrupting body has  yet to be quantitatively modeled, making a comparison with P5 difficult.   This is, in part, because the appearance is likely to be dominated by small particles that carry most of the cross-section of ejected material while most of the mass resides in large particles which precipitate the instability.  Other model problems relate to uncertainties in the mechanical properties and in the basic physics of disintegration of rotating aggregate bodies.  A qualitative expectation is that rotational ejection would release low velocity ejecta (speeds comparable to the 0.3 m s$^{-1}$ escape speed of the nucleus), with no fast ejecta.  This is broadly consistent with the success of synchrone-fits (which assume zero initial velocity).  Rotational mass-shedding could be intermittent as unstable clumps of material migrate towards the rotational equator, break and detach from the central body.   The escaping material should be largely confined to the plane of the equator of the central body, as reported for fragments in another MBC, P/2010 A2 \citep{2013ApJ...769...46A}, whose orbit is very similar to that of P5 and which may also be a Flora family member.     On the other hand, the morphologies of these two objects are quite different, so it is not obvious that they share a common origin.


\clearpage

\section{Summary}

The main properties of active asteroid P/2013 P5 determined from Hubble Space Telescope observations are summarized as follows:

\begin{enumerate}

\item The object shows a central nucleus (absolute magnitude $H_V \sim$ 18.6, corresponding to a mean radius of 0.24$\pm$0.04 km or less, with assumed geometric albedo $p_V$ = 0.29$\pm$0.09) embedded in a system of six, divergent dust tails.  The scattering cross-section of the dust exceeds that of the nucleus by a factor $\sim$13.

\item There is dramatic morphological change in the tails between UT 2013 Sep 10 and 23, but very little photometric change in the nucleus or near-nucleus dust environments.   

\item The position angle of each tail can be characterized by a different date of ejection, with a five month age span (from UT 2013 April 15 to Sep 04) that indicates continuing activity at the nucleus.  

\item Mass loss through rotational disruption is the most plausible mechanism driving the mass loss.

\end{enumerate}

\acknowledgments
Based on observations made with the NASA/ESA \emph{Hubble Space Telescope,} with data obtained at the 
Space Telescope Science Institute (STScI).  Support for program 13475 was provided by NASA through a grant from the Space Telescope Science Institute, which is operated by the Association of Universities for Research in Astronomy, Inc., under NASA contract NAS 5-26555.  We thank Alison Vick, Tomas Dahlen, and other members of the STScI ground system team for their expert help in planning and scheduling these Target of Opportunity observations.

\clearpage

\begin{deluxetable}{lccccccc}
\tablecaption{Observing Geometry 
\label{geometry}}
\tablewidth{0pt}
\tablehead{
\colhead{UT Date and Time}  & \colhead{$R$\tablenotemark{a}}  & \colhead{$\Delta$\tablenotemark{b}} & \colhead{$\alpha$\tablenotemark{c}}   & \colhead{$\theta_{\odot}$\tablenotemark{d}} &   \colhead{$\theta_{-v}$\tablenotemark{e}}  & \colhead{$\delta_{\oplus}$\tablenotemark{f}}   }
\startdata
2013 Sep  10 16:44 - 17:24 & 2.112 & 1.115  & 5.1 & 125.0 & 244.8  & -4.2\\
2013 Sep  23 09:20 - 09:59 & 2.096 & 1.135  & 10.7 & 89.2 & 244.5  & -4.3\\

\enddata


\tablenotetext{a}{Heliocentric distance, in AU}
\tablenotetext{b}{Geocentric distance, in AU}
\tablenotetext{c}{Phase angle, in degrees}
\tablenotetext{d}{Position angle of the projected anti-Solar direction, in degrees}
\tablenotetext{e}{Position angle of the projected negative heliocentric velocity vector, in degrees}
\tablenotetext{f}{Angle of Earth above the orbital plane, in degrees}

\end{deluxetable}

\clearpage

\begin{deluxetable}{lccccc}
\tablecaption{Photometry\tablenotemark{a} 
\label{photometry}}
\tablewidth{0pt}
\tablehead{
\colhead{Date}      & \colhead{$V_{0.2}$\tablenotemark{a}} & \colhead{$V_{1.0}$\tablenotemark{b}} & \colhead{$V_{4.0}$\tablenotemark{c}}  & \colhead{$V_{6.0}$\tablenotemark{d}} & \colhead{$H_V$\tablenotemark{e}} }
\startdata
2013 Sep 10     & 20.92  &20.47  & 18.46  & 18.38 & 18.69 \\
2013 Sep 23     & 21.01 & 20.57  & 18.41 & 18.34  & 18.54  \\
\enddata


\tablenotetext{a}{All magnitudes have a statistical uncertainty of $\pm$0.01 mag.~or better, but systematic uncertainties which grow with aperture radius and are at least several $\times$0.1 mag.~in the largest aperture.}

\tablenotetext{a}{Apparent V magnitude within 5 pixel (0.2\arcsec) radius aperture}
\tablenotetext{b}{Apparent V magnitude within 25 pixel (1.0\arcsec) radius aperture}
\tablenotetext{c}{Apparent V magnitude within 100 pixel (4.0\arcsec) radius aperture}
\tablenotetext{d}{Apparent V magnitude within 150 pixel (6.0\arcsec) radius aperture}
\tablenotetext{e}{Absolute V magnitude computed from $V_{0.2}$ using Equation (\ref{inversesq})}

\end{deluxetable}

\clearpage

\begin{deluxetable}{ccccrr}
\tablecaption{Tail Position Angles 
\label{angles}}
\tablewidth{0pt}
\tablehead{
\colhead{Feature\tablenotemark{a}}  & \colhead{Sep 10\tablenotemark{b}}    & \colhead{Sep 23\tablenotemark{c}}   & \colhead{DOY\tablenotemark{d}} & \colhead{Date\tablenotemark{e}} & \colhead{Age\tablenotemark{f}} }
\startdata
A & 237 $\pm$ 1 & 234 $\pm$ 1 & 105  & Apr 15  & 161 \\
B & 220 $\pm$ 1 & 198 $\pm$ 1 & 199  & Jul 18  & 67  \\
C & 216 $\pm$ 2 & 190 $\pm$ 2 & 205  & Jul 24  & 61 \\
D & 202 $\pm$ 1 & 153 $\pm$ 2 & 220 & Aug 08  & 46 \\
E & 161 $\pm$ 2 & 114 $\pm$ 1 & 238 & Aug 26  & 28 \\
F & 141 $\pm$ 2 & 97 $\pm$ 2 & 247  & Sep 04  & 19 \\
\enddata


\tablenotetext{a}{See Figure (\ref{image})}
\tablenotetext{b}{Position angle on UT 2013 Sep 10, in degrees}
\tablenotetext{c}{Position angle on UT 2013 Sep 23, in degrees}
\tablenotetext{d}{Day of year of best-fitting synchrone ejection (uncertainty is $\lesssim$1 day)}
\tablenotetext{e}{Calendar date corresponding to DOY}
\tablenotetext{f}{Tail age on Sep 23, in days}
\end{deluxetable}

\clearpage

\clearpage

\end{document}